Elizabeth Locci (elizabeth.locci@cern.ch)     31/08/2015

# DEVELOPMENT OF WIRELESS TECHNIQUES IN DATA AND POWER TRANSMISSION APPLICATION FOR PARTICLE-PHYSICS DETECTORS

## - WADAPT -
## Wireless Allowing Data And Power Transmission

**Proposal**

R. Brenner[a], S. Ceuterickx[b], C. Dehos[c], P. De Lurgio[d], Z. Djurcic[d], G. Drake[d], J.L. Gonzalez Gimenez[c], L. Gustafsson[a], D.W. Kim[e], E. Locci[f], D. Röhrich[g], A. Schöning[h], A. Siligaris[c], H.K. Soltveit[h], K. Ullaland[g], P. Vincent[c], D. Wiednert[h], S. Yang[g]

[a] *Uppsala University, Sweden*
[b] *CERN, European Organisation for Nuclear Research, Genève, Switzerland*
[c] *CEA/LETI/DRT/DACLE/LAIR, Grenoble, France*
[d] *Argonne National Laboratory, Argonne, IL 60439, USA*
[e] *Gangneung National University, Korea*
[f] *CEA/DSM/IRFU/SPP, Gif-sur-Yvette, France*
[g] *University of Bergen, Norway*
[h] *University of Heidelberg, Germany*

## Abstract

Wireless techniques have developed extremely fast over the last decade and using them for data and power transmission in particle physics detectors is not science-fiction any more. During the last years several research groups have independently thought of making it a reality. Wireless techniques became a mature field for research and new developments might have impact on future particle physics experiments.
The *Instrumentation Frontier* was set up as a part of the *SnowMass 2013 Community Summer Study* [1] to examine the instrumentation R&D for the particle physics research over the coming decades: « To succeed we need to make technical and scientific innovation a priority in the field ».
Wireless data transmission was identified as one of the innovations that could revolutionize the transmission of data out of the detector. Power delivery was another challenge mentioned in the same report.

We propose a collaboration to identify the specific needs of different projects that might benefit from wireless techniques. The objective is to provide a common platform for research and development in order to optimize effectiveness and cost, with the aim of designing and testing wireless demonstrators for large instrumentation systems.

**Table of contents**

# 1. Introduction

This document describes the motivations for and the state of wireless instrumentation techniques, discusses present developments and describes an approach to form a collaboration with the goal to design and develop demonstrators for a large range of particle-physics detectors. Possible wireless applications encompass high-energy-physics (HEP) experiments at high-energy colliders, neutrino-physics experiments, astroparticle-physics experiments and low energy experiments at the intensity frontier. Considering the large domain of applications and experimental conditions, the motivations and requirements, with more or less stringent challenges, are examined. The present status of research and the commercially available components are explored. The use of wireless techniques for data and power transmission is exemplified by two recent developments, in neutrino physics and in HEP, that operate under very different conditions. Given the large domain of applications, in terms of detector specificities, experimental environment and other constraints, the work would profit from a well-organised and focused collaboration that helps in finding common solutions to common problems and in identifying specific developments.

# 2. Motivation

The motivation for wireless power and data transmission is manifold in particle physics detectors, and among them:

- Wireless technologies offer a unique and very elegant opportunity to send broadcasts. This is particularly interesting for steering and control of a complex detector system and might save a lot of cables if one single signal is sent to many receivers.

- The total or even partial removal of cables and connectors will result in cost reductions, simplified installation and repair, and reductions in detector dead material. These two last aspects are especially important in tracking detectors and may become particularly important in case of limited access or/and hostile environment.

- Wireless data transfer offers the possibility to realize topologies which are much more difficult to be realized using wires, as data from one single point can be sent to several receivers or several transceivers send to one receiver.

These features may become of particular importance for future HEP detectors.

- The increasing demand for high data transfer rates from highly granular detectors in HEP that is limited today by the available bandwidth of electrical and optical links. The weaknesses of both these technologies are the size of the connectors and the sensitivity of the link to mechanical damage. The proposed wireless data transmission, using the 60 GHz or higher band, offer in this context the required bandwidth, high space efficiency, security and form factor.

Minimizing the amount of material in the region of the tracking detectors will reduce multiple scattering and nuclear interactions that degrade the precision on the measurement of track momentum and interaction vertices, and in addition will reduce the number of fake hits arising from secondaries. Well-chosen detector technology and geometry, combined with wireless techniques might help in reducing the amount of cables and optimizing their path, and thus minimizing the dead-zone areas.
Fast triggering, as exemplified in section 4, using all information from the tracker, and possibly from other detectors, is essential for hadron colliders at high luminosity. Although less mandatory for lepton colliders, introducing wireless-links intelligence providing communication between sections (tracking layers, for instance) of a detector within a region of interest might be useful [2-7].

## 3. Wireless Technology

### 3.1 Data Transmission

Wireless information transmission (WIT) is far from being a new idea. Optical transmissions, in their most simple form, were used more than 2000 years ago and revisited by Bell and Tainter who invented and patented the photophone in 1880. Simultaneously radio communication started to develop from the demonstration of the existence of electromagnetic waves by Hertz in 1888. In 1896 Marconi developed the first radio-telegraph system and public use of radio began in 1907, with many great contributors since then. In 1947/1948 the concept of cellular technology was developed and the basis for data compression was shed. Packet technology started to develop from the 1960's, and in the 1980's several cellular radio networks were deployed around the world. Growth in commercial wireless technologies basically started in the 1990's with the use of bands of ~ 1 KHz and with data rates of approximately 50 Kbps to 2 Mbps at the dawn of the 2000's.

The original version of the standard IEEE 802.11, well known as Wi-Fi, was released in 1997 and delivers about 1 Mbps of payload traffic. A decade later, the standard 802.11n that is used for one of the two applications described in section 4.1 is able to deliver a payload throughput of more than 200 Mbps. Latest technology development such as the 802.11ac and 802.11ad standard (December 2012) are respectively able to deliver 800 Mbps on the 5 GHz UNII band and 3.2 Gbps on the 60 GHz ISM band. Other wireless technologies such as satellite communications, cellular networks or dedicated point-to-point wireless communication system know the same evolution.

During the last decade performances in terms of data rate have constantly increased from less than a few tens of Mbps (IEEE 802.11n, 3GPP UMTS (3G cellular), IEEE 802.16 (WiMAX)) to a few Gbps (IEEE 802.11ad and 802.11ac, 3GPP LTE-B (5G cellular)). Significant improvements were also made in radio link reliability, medium access, power consumption and system cost.

For years, radio communication technology had tried to reach the channel capacity limit as defined by the Shannon-Hartley theorem without exceeding it. The theoretical physical data rate at which information can be transmitted over a communication channel is a function of the bandwidth and the signal-to-noise ratio[1]. The recent development of the MIMO (Multiple-Input, Multiple-Output) radio systems has enabled to overtake this limit. Making use of the spatial dimension of a communication link, the maximum channel capacity is still a function of bandwidth and signal-to-noise ratio but it is increased by a factor n of the number of spatial streams.

The fast development of the wireless technologies is linked to the improvements made in the electronics, especially the semiconductor devices and the new simulation capacity [8-14]. Some of the main strategies to improve the performance of radio

---

[1] here the signal to noise ratio is defined as the average received signal over the bandwith, and the noise is defined as the average noise or interference power over the bandwidth

communication systems are the following:

- Over the last decade the spectral efficiency, meaning the data rate of the physical layer that could be transmitted in a given radio frequency bandwidth, has been increased by a factor 10. This was achieved by the use of more efficient transmission techniques such as spread spectrum, UWB or OFDM and higher-order modulation such as QPSK, QAM. The introduction of smart antenna designs has not only increased the channel capacity with the MIMO technology but also improved the radio link quality under conditions of interference and signal fading (eg. Multipath). The use of these advanced techniques requires intensive signal processing, increasing the complexity of the chipset design, the size as well as the power consumption. These constraints are now properly tackled by the industry, which has been able to increase of the embedded computing power in the radio chipset while containing the power consumption, heat dissipation and manufacturing cost.

- According to the Shannon-Hartley theorem, another strategy to increase the throughput is to increase the spectrum bandwidth. However, larger bandwidths require transmission at higher frequency, which became feasible, once again, with the improvement made on the electronics. For example, the IEEE 802.11ad running in the 60 GHz band allows up to 9 GHz of total bandwidth while the IEEE 802.11ac running in the 5 GHz band benefits from less than 500 MHz of bandwidth.

- According to the Free Space Path Loss equation, the attenuation of a 60 GHz transmission at 1 meter is about 68 dB, which is 21.6 dB higher than a 5 GHz transmission. Consequently at constant transmit power and using modulation scheme, the communication range will be 12 times shorter. This shorter signal range could be either a challenge to setup high throughput link over several meters or a benefit to mitigate co-channel interference making a high-density channel-reuse design possible.

- The higher frequencies lead to smaller sizes of the RF components, including antennas, which can be easily integrated into electronic systems. The application-related antenna must be carefully chosen according to many parameters such as directionality, bandwidth, gain, etc. A wide range of antennas can be produced with standard methods used for electronics and printed circuits boards. The 60 GHz band enable the use of Manufacturing techniques such as Low-Temperature Co-fired Ceramic, silicon, organic material to build on the printed circuit board close the radio module small size and high gain directional antennas. The use of RF lenses to increase the gain and the directivity of the antenna is also a possibility at this frequency.

- Improvements lie in carefully selecting the communication protocol. For a radio communication channel to work, a large amount of the transmission time is devoted to transmitter and receiver synchronisation, medium access, medium sensing, legacy standard-protection mechanism or transmission error compensation. In a 802.11 communication system for example, about half of raw throughput is "lost". Latest standards have introduced new mechanisms such as SDMA and frame aggregation to improve the efficiency. In a particle physics detector the propagation environment could be well defined and controlled allowing a reduction of the overhead in the benefit of the payload traffic and a significant improvement of the jitter and latency. The distribution of a system clock in particle physics detectors will facilitate the transmitter-receiver synchronisation.

Supported by market demand for even more throughput and even more reliability, these wireless technologies continue to develop with an incredible speed (Figure 1). Setting up multiple neighbouring gigabit wireless links is today a reality for home, industrial and business applications. Higher modulation rates such as 1024QAM are already deployed for some point-to-point links and 4096QAM have been achieved in the laboratory. Developments are also ongoing in the smart antenna field [15-23] promising a better control on the radio link quality and sharing. The 240 GHz [24] or 280 GHz [25] technologies for high data rate (40 to 100 Gbps) are on the rails as prototypes, and might become industrialised within three to five years.

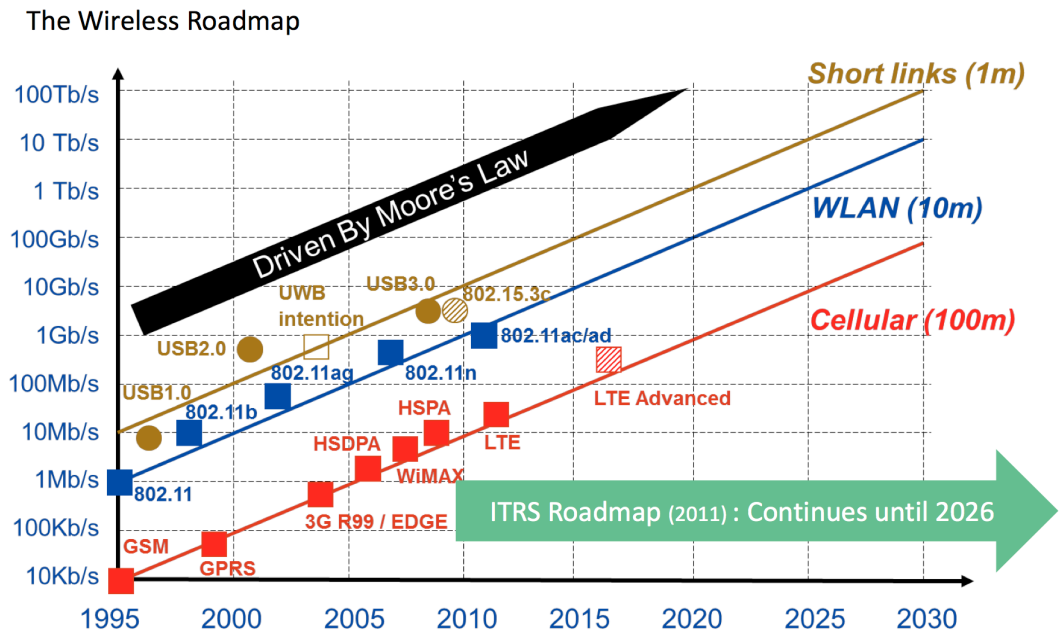

Figure 1: ITRS Wireless Roadmap [26-28]

To conclude, the latest research performed on Orbital Angular Momentum (OAM) multiplexing technique [29] seems very promising and could be the next breakthrough to reach Terabit wireless transmission.

**3.2 Power transmission**

The concept of Wireless Power Transmission (WPT) is not new either. In the late 19$^{th}$ century, both Herz and Tesla theorized the possibility of wireless power transmission and Tesla demonstrated the feasibility. Efforts to use this technology developed during World War II, and as a result of these efforts, in 1964, the televised demonstration of a microwave-powered helicopter broadly introduced the concept of wireless data transmission to scientific and engineering communities and to the public [30]. In 1982, Brown developed a low-weight rectifying antenna (rectenna) that led to the development of the Stationary High Altitude Relay Platform (SHARP). The most documented area of research is being done with the goal of putting Solar Power generating Satellites (SPS) into space and transmitting power to the Earth. The idea was proposed by Glaser in 1968, but so far experiments have only been carried out in terrestrial laboratories. However a Japanese government agency is planning the construction of a 1 GW commercial SPS in 2030's [31]. In the late 2000's, WPT started being adopted for medical use (pacemakers, etc..) and commercial use (laptops, cell phones, tablets, etc..).

The power efficiency is the critical criterion for WPT and determines the choice among the various technologies: inductive coupling, magnetic resonant coupling, electromagnetic radiation.

Under inductive coupling the power transfer falls off steeply even over a very short distance. Compared to inductive coupling, magnetic resonant coupling can achieve higher transfer rate efficiency over larger distances (~ 2 m). But this is still insufficient to be considered for large detectors in HEP. RF-based WPT might be the optimal choice in terms of simplicity and cost. Since radio signal can carry energy as well as information at the same time, Simultaneous Wireless Information and Power Transfer (SWIPT) would be an appealing concept. However the present designs are far from reaching our goals in terms of data rate versus received power.

More understanding of possible fundamental limitations is needed. The chosen technology is extremely dependent on the application; for instance RF in the frequency band around 2 or 5 GHz and below would strongly affect the electronics that is very sensitive to noise in this frequency band.

## 4. Experimental Context

### 4.1 Data Transmission

The application of wireless techniques has been studied in two different contexts:

**Neutrino Experiments**

The primary purpose of the effort described in reference [32] is to ascertain the feasibility and practicality of using wireless techniques for a large detector containing photomultiplier tubes (PMT). The choice of PMTs is partly guided by the ubiquity of PMTs in HEP detectors. The capability of measuring single photoelectrons is a common demand, which imposes requirements on bandwidth and sensitivity of the instrumentation. Such requirements are often listed in connection with large neutrino detectors. Some examples include monitoring of the neutrino flux at a nuclear reactor, large mobile neutron detectors for detection of radioactive materials in security applications, and detectors operating in high radiation areas.

The design assumes 6 Bytes/event (2 for the pulse height, 4 for the time-stamp), with a maximum event rate (single p.e.) of 10 kHz, which translates to 480 kbps for a single channel. Extrapolating to a large detector of 50000 channels translates into a data rate of 24 Gbps.

Different technologies may be considered for wireless transmission, however RF transmission was preferred to optical transmission as it does not require line-of-sight, and an individual transmitter can communicate with many front-ends (point-to-multipoint communication). WLAN (Wireless Local Area Network) technologies offer higher data throughput than mobile/cellular technologies, and among them 802.11n was the most suitable at that time. The total data rate of a single 802.11n 20MHz stream is ~ 65 Mbps, with a payload data rate of ~ 35 Mbps. One frequency range in 802.11n is centred at about 5.5 GHz, with an overall bandwidth of ~ 1.2 GHz. Single stream 802.11n access points can have an individual operating bandwidth of 20 MHz or 40 MHz. 20 MHz wide channels can accommodate 48 access points. For a single stream 802.11n link this translates into a total payload data rate of 1.68 Gbps, as illustrated in Figure 2. For large detectors with tens of thousands or millions of channels, this payload data rate is clearly insufficient. This data throughput can be increased in two ways. The use of directional antenna and the intrinsic shielding created by the detector would allow the same channel to be reused in some different parts of the detector. The use of up to 8 MIMO (multiple input and output) streams would translate into a much higher payload data rate per stream of 7.68 Gbps to be compared to 1.68 Gbps.

A prototype (Figure 3) was built targeting a 35 Mbps transfer rate, with an error rate less than $10^{-12}$. The achieved performance was 11 Mbps, which could be increased to 18 Mbps by reducing the latency with a simple change. No bit error was observed in any received packet, however a significant number of packets were dropped (1/2000), and this was attributed to the chosen protocol.

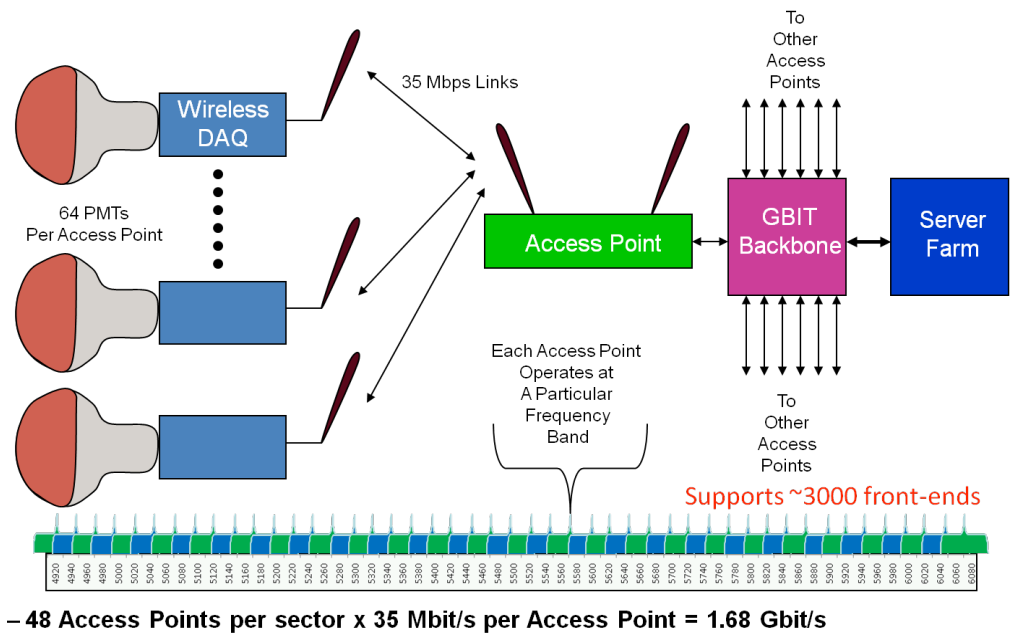

Figure 2: Allocation of frequency space for wireless data transmission

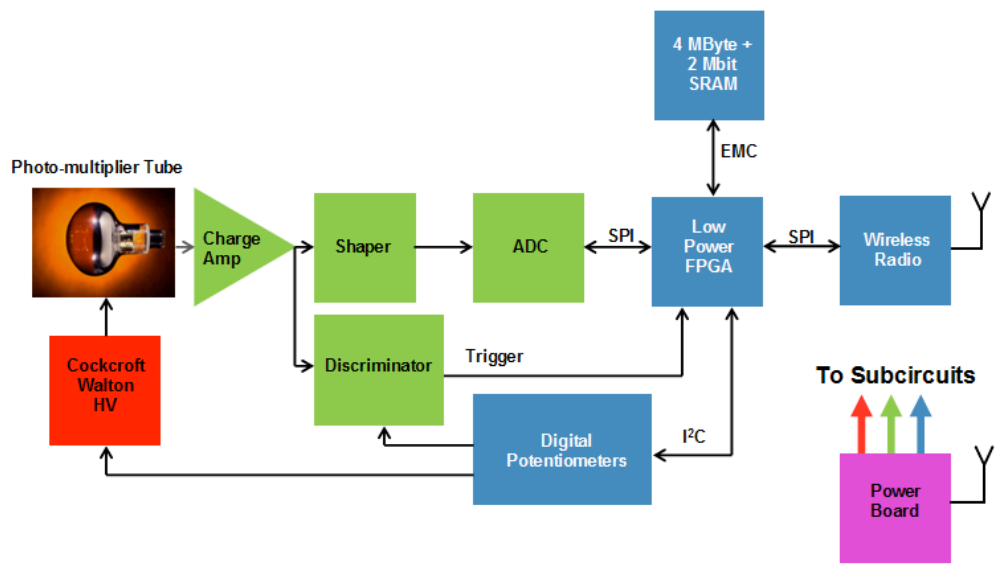

Figure 3: Block diagram of the wireless prototype system

802.11 standards are making use of various techniques (CSMA/CA, protection mechanism, data rate adaptation, frame acknowledgement etc) which impact the overall performance; throughput but also packet loss, jitter or latency. The purpose of these is to enable the medium sharing and maintain a reliable connectivity when the link quality decreases. The direct impact is that the payload data rate for 802.11 connection is, in the best case, half of the physical data rate. However, in the specific application of a detector, it could be assumed that the wireless link quality and the medium would be known and controlled. This would allow implementing a lighter protocol, which would provide lower overhead and predictable performance.

**LHC Experiments**

In the last decade there has been tremendous advances in silicon technologies that have made it possible to build high performance transceivers operating in the millimetre-wave band, where the 57-66 GHz band is situated. This license free 9 GHz band is very attractive in order to achieve high data rate transfer that has triggered the use in the HEP environment.

In addition to the high data rates that is possible using the 60 GHz spectrum, the energy propagation characteristic of this band is unique. It has a free space path loss of 68 dB over a distance of 1 m, a high material penetration loss, that in our case is measured to be about -50 dB (preliminary value) for a fully equipped ATLAS Semiconductor-Tracker detector module, and an oxygen absorption of about 15 dB/km. The last effect is of less importance for us, since a typical data transmission distance in HEP detectors is from a few cm to about 1 m for which an attenuation of about 0.1 dB is expected.

Antennas operating at such high frequency are typically very directional, unlike that operating at 2 or 5 GHz. Directivity is a measure of how well an antenna focuses its energy in an intended direction, thus operating at 60 GHz frequency results in a more focused antenna with a narrower beam width for a fixed size antenna, that minimizes the possibility of interference and the risk that the transmission be intercepted. These features, the high path loss, high material penetration loss, narrow beamwidth, Line-Of-Sight (LOS), and operation in a controlled environment, makes the 60 GHz band optimal for short range operation. Also the use of the high carrier frequency provides low form factor, which will reduce the material budget. This provides an extremely desirable frequency re-use that can handle a large number of transceivers in a small area as in the HEP detectors and other detectors facilities.

The work described in reference [6,7] aims at demonstrating the feasibility of wireless readout of the Silicon inner-tracker for the ATLAS silicon strip detector with use of the 60 GHz band (Figure 4).

The 60 GHz band is very suitable for high data rate and short distance applications, which can provide wireless Multi Gigabit per second radial data transmission inside the ATLAS silicon strip detector, making a first level track trigger processing all hit data feasible.

For a complete readout of the silicon micro-strip tracker a bandwidth of about 50-100 Tbps is required. So with 20,000 links, a bandwidth of 5 Gbps per link would be required. The use of the 60 GHz band associated to a large spectral bandwidth (9 GHz) would make a few 10's of Gbps achievable. The block diagram of the proposed 60 GHz transceiver chain is illustrated in Figure 5.

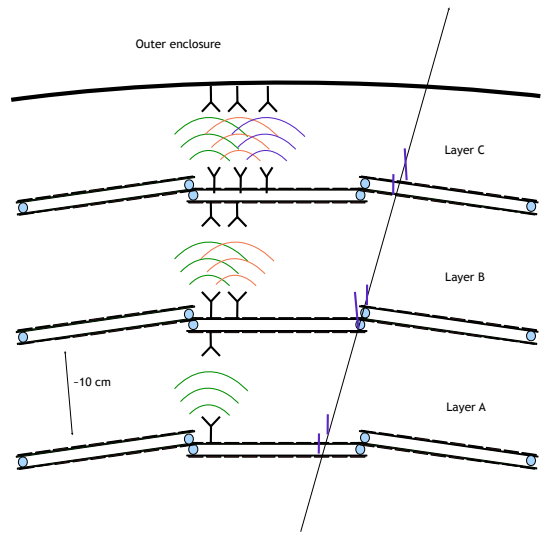

Figure 4: Proposal of a radial readout for the tracker detector of the ATLAS experiment

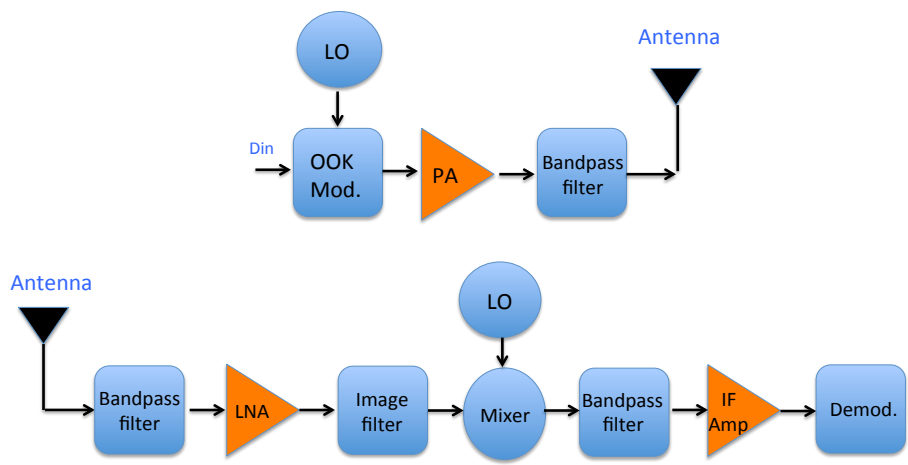

Figure 5: Block diagram of the transceiver. The transmitter is shown at the top and the receiver at the bottom.

The first prototype is designed to handle a data rate of 4.5 Gbps over a link distance of 1 m. Estimated power consumption for a first full prototype readout is less than 240 mW. The chosen technology must be able to fulfil requirements such as noise and linearity, radiation hardness, and at the same time have a high production yield at a reasonable cost. The 130 nm SiGe Bi-CMOS HBT 8HP technology has been chosen for the first prototype. Radiation hardness at increased radiation level as expected for the ATLAS upgrade, for instance, needs to be tested, and some radiation hard layout for such conditions may be necessary.

The for the first prototype chosen On-Off Keying (OOK) modulation has the benefit of simplicity but the drawback of low spectral efficiency (0.5 bps/Hz) and high noise sensitivity. More advanced modulation and transmission techniques like OFDM and MIMO could be investigated. For example, the spectral efficiency of 802.11ad wireless standard is about 3 bps/Hz for the highest modulation (OFDM 64QAM 13/16). The throughput at the hardware layer would be increased by a factor six. In addition, the use of antenna diversity or MIMO techniques could also enhance the link reliability. These techniques are commonly used in wireless communication to mitigate the impact of wave reflections. Moreover, MIMO has the benefit to significantly increase throughput without additional spectral bandwidth by making use of the wave reflection issue. Although these techniques are today mature technologies, the design would be more complex and the power consumption will increase.

## 4.2 Power Transmission

Reference [32] not only describes the design and testing of wireless data transmission but also that of wireless power transmission. Both optical and RF power transfer methods were tested.

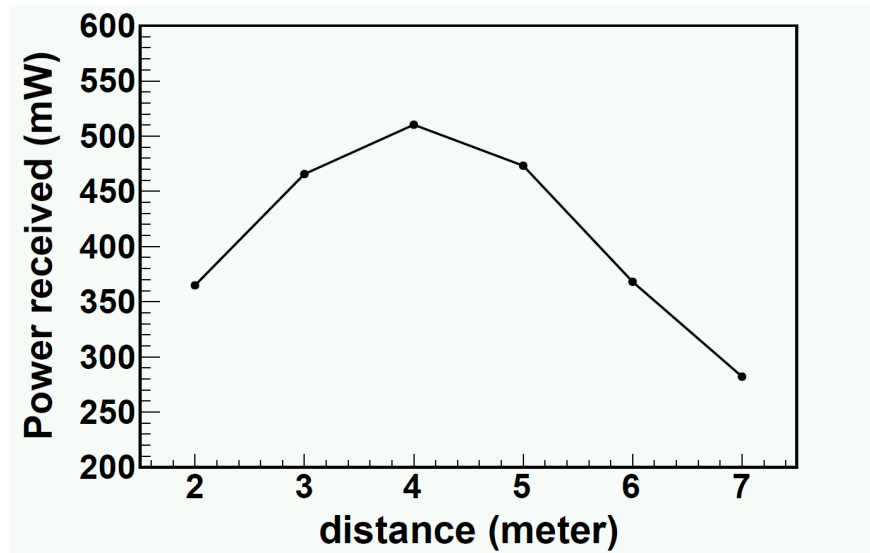

Figure 6: Power received (mW) by the photovoltaic panel as a function of distance (m) from the optical source. The optical source used was a 3.5 W LED with a 940 nm wave-length and a focusing lens.

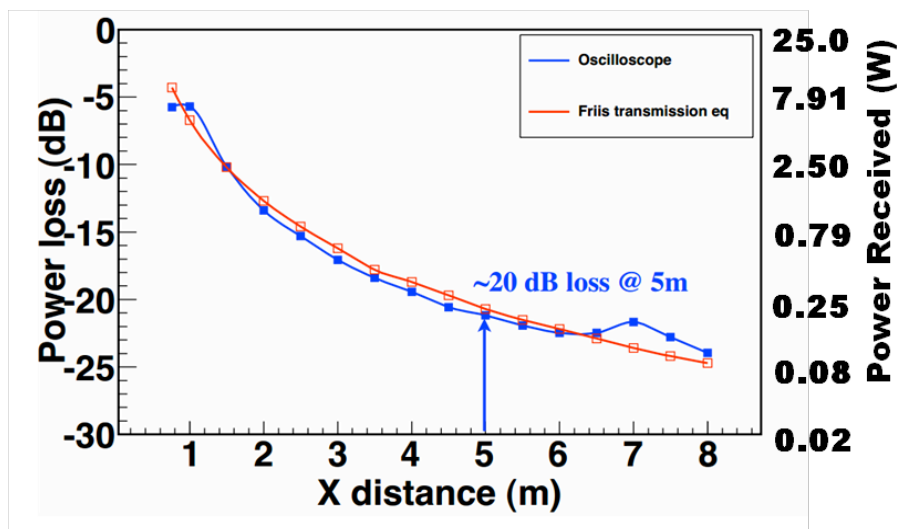

Figure 7: Power loss (dB) in the RF power transmission as a function of distance (m) from the source. Solid dots are the measurements and open dots are the expectation from Friis transmission equation. The source is a 915 MHz 10 dBm signal transmitted through a 14 dBi gain Yagi antenna and received by an 11 dBi gain patch antenna. The scale on the right shows the power that would be received for a 25 W (44 dBm) source.

The optical demonstrator utilizes a high power light-emitting diode, with 3.5 W optical output, collimated and received by a photovoltaic panel. The received power depends on the distance between the source and the panel, with a maximum of 500 mW at ~ 4 m and then decreases to ~ 280 mW at 7 m (Figure 6).

The RF power transfer test consists of a function generator driving a 14 dBi gain Yagi antenna at 915 MHz with an output power of 10 dBm and being received by a 11 dBi gain patch antenna. There is a rapid fall off of the received power as the transmission distance increases: from ~ -5 dBm at 1 m to ~ -23 dBm at 7 m (Figure 7). The prototype requirements of receiving 250 mW at 5 meters was easily met by the optical device whilst a 25W source would be needed for the RF device.
Nevertheless RF power, making use of highly directive antennas, would be a better choice for large detectors since one source can power many front-ends.

## 5. Objectives and Strategy

An all-wireless large detector for HEP is an elegant and attractive concept in itself. Some motivations have been illustrated by the examples of quoted developments: elimination of massive cable plants, fast data transfer for first-level trigger. The elimination of massive cable plants could result in possible cost reductions, simplified installation and repair, and reduction in dead material, all being aspects that might be important in harsh environments.

The first objective is the **precise definition of the common and specific needs of particle-physics detectors** which depend on the experimental context (LHC, FCC-ee, FCC-pp, ILC, CLIC, others) and the chosen detector technologies:

- Topology
- Number of channels per detector (set of channels)
- Power required per channel in a given set
- Expected data rate per channel in a given set
- Local environment (implementation, radiated heat, radiation, magnetic field, reflections, attenuations, couplings, etc.)
- Global environment (experimental setup, machine operation mode, etc.)

The second objective is the **evaluation of the present technologies for data and power transfer**:

- For **data transfer,** the 5.5 GHz technology is well established and well commercialized, and its maximum payload data rate of about 1.7 Gbps (160 MHz bandwidth, 4 streams, higher modulation rate) makes it suitable for a few thousands channels in the telecommunication domain, although it may be over-optimistic for particle physics applications. For larger numbers of channels, the 60 GHz technology, with a physical data rate of ~ 6 Gbps per channel per stream, is better adapted and is commercially available. The 240/280 GHz has demonstrated being able to transmit 40 Gbps, and might be able to reach 100 Gb/s, but is currently at the level of the prototype and a couple of years is expected to be necessary to reach the level of industrialisation. Such high rates are currently achievable with optical links. The weaknesses of optical links are the size of the connectors and the sensitivity to mechanical and radiation damage. The advantage of wireless transmission over optical transmission is quite clear for the application described in Reference [32]. The advantages/disadvantages of different antenna types (horn, patch, lens …) antennas have to be studied. For a general use of wireless data transmission, the advantages have to be evaluated in terms of implementation, material budget, cost and long term reliability in harsh environment.

- **Power transfer** is more complex. The main advantage would be the total elimination of cables, mostly in remote or harsh environments, however is this practical? The major limitation of the WPT is the efficiency that is strongly dependent on the experimental context, and more particularly on the distance between the transmitter and the receiver. High efficiencies (70 to 90 %) can be obtained with inductive or resonant magnetic coupling on distances comparable to the size of the coil; these techniques are quite suitable for medical applications but may not be appropriate for HEP experiments. The RF technique could be considered, however its efficiency is much lower (~ 30 % at 1 m to ~ 1% at 5 m). The efficiency, depending on the gain of the antennas and on the selected wavelength, may be optimized by tuning these parameters. As for the WIT the advantages of the WPT have to be assessed in terms of implantation, material budget, cost.

For most applications in particle physics, the environment is much different from traditional wireless communication environment, such as office buildings or open air. The chosen technologies will be required to operate under harsh conditions due to:

- the strong magnetic field (typically 3 to 6 Tesla) in some cases
- the radiation created by the beam in some cases
- the RF noise/interference created by neighbour cells of the wireless system itself
- the signal crosstalk and multi-path propagation on metallic parts of the detector
- the variations of temperature and humidity that might impact the electronics

The chosen wireless technologies will have to sustain these conditions with a high level of reliability over 10 to 20 years.

The commercially available chipsets are not designed to work over 20 years under those stringent conditions. However some are designed to support intense weather conditions: 40 to 80°C with 100% humidity and some others are qualified to run in the aerospace industry. Usually the consumer premise chipsets are not calibrated causing large fluctuations in the RF properties; dispersions exceeding 6 dB are common even with some professional chipsets. Moreover part of the layer 1 and 2 (OSI model) of commercial chipsets is locked, making any change in the working protocol impossible or limited at the hardware level.

The monitoring and troubleshooting inside a detector is complex; how to bring a spectrum/network analyser inside a detector layer is a question to be answered. However, the industry, in particular the cellular and satellite carriers, continuously develops mechanisms to remotely and automatically diagnose, repair and improve the availability and performance of their wireless systems. Some dual radio communication systems are been designed to perform loopback testing or with embedded "light" spectrum analyser in order to report back link quality assessment.

Maintaining a common time reference throughout a wireless sensor network (WSN) is an essential requirement for coordinated measurements over a large detector. Any error affecting the timestamp has a direct impact on the final synchronisation accuracy of the system and thus on the efficiency of event reconstruction. A precise time-stamping network has been shown to be feasible [33] by the use of dedicated signal-process algorithms that can be implemented within each node wireless receiver.

Several questions have to be answered: Are commercial chipsets suitable for particle-physics detectors? Can an actual advantage of wireless transmissions over wired ones be demonstrated? What about the long term performance and support?

Operating wireless technologies in large detectors is a challenge that will require years of R&D to design, adapt or build, evaluate and test a full radio communication system (antenna, chipsets and communication protocol) for our applications.

Once the needs have been clearly identified and the technology availability and practicability have been assessed, the third objective would be the **simulation** of concrete cases, as the described ATLAS project for HEP, for instance. With a huge number of radiating systems that can interact with each other, it is mandatory to estimate the couplings with electromagnetic simulations on models of high geometric fidelity. Accurate model geometries can be built by importing CAD files or by using some geometry-building systems. Some commercial software is available to perform this study and to establish standards for the in-situ installation that respect EMI/EMC immunity guidelines.

The tools that would be developed should be easily adaptable to the particularities of other experiments (FCC-ee, FCC-pp, ILC, CLIC, others). RF simulation of multi-antenna systems need to be developed. Detector-layout may be affected by the LOS and thus must be integrated in simulation studies.

On the basis of simulations, some small-scale **prototypes** can be constructed and tested, in order **to establish the feasibility of WIT and WPT for large detectors**, which is our fourth objective.

The strategy to reach these objectives implies a close contact with developers of the WIT and WPT technologies and with industry in order to obtain ultimate performances and reduced costs. It also requires a strong effort on simulation and tests.

## 6. Work Plan: Time Scale, Working Groups, Milestones

The two projects given as examples are well-defined and advanced projects, spanning over 3 to 5 years, which can be the nucleus of the future R&D Collaboration, that may grow as more people come with precise projects. The global project of R&D would cover the next decade and might be extended for another decade depending on the time scale of the future experiments.

A Collaboration Board (CB) would complete/modify the proposed Working Group (WG) structure. Each working group will precisely define its program of R&D that will be reviewed by the CB.

The main WGs would be:

- Wireless Information Transmission (WIT) that would coordinate all activities related to WIT

- Wireless Power Transmission (WPT) that would coordinate all activities related to WPT

- Wireless Technology Test (WTT) that would coordinate follow-up of new developments in wireless technologies, tests of components, tests of prototypes, construction of demonstrators.

- Wireless Simulation Tools (WST) that would coordinate all activities involving local as well as global simulations.

- Wireless Standard Definition (WSD) that would work in close contact with the WST working group to elaborate rules for the implantation of the wireless technology in the detectors and would control that the implantation strictly follows these rules.

WIT and WPT working groups will necessarily strongly interact with the three other working groups WTT, WST and WSD.

A first natural milestone would be the completion of the two projects given as example in a time scale of a few years. Then the status of the global project should be reviewed, and more precise milestones could be defined depending on this status and on the plans for future experiments.

## 7. Expertise

In the few last years, two groups have independently thought of using wireless techniques in the domain of detectors for particle physics. As described in section 4, two experimental projects have started developing in parallel:

- In the context of an HEP experiment, ATLAS at the LHC, groups from the Heidelberg and Uppsala Universities are working in close contact on the wireless readout of the silicon strip detector. Both groups are formed from experienced senior physicists and engineers who gained expertise from their work on large experiments such as H1 at HERA, DELPHI, OPAL at LEP, ATLAS, ALICE and LHCb at LHC as well as in RD collaborations such as RD20. Under the supervision of the Heidelberg group a student has completed a thesis on the development of a test setup for the 60 GHz wireless transceiver for the ATLAS tracker readout, and is actively working on the test facilities. From the ALICE group, the University of Bergen has shown great interest in the project for the LS3 upgrade; three students are already assigned for the project development. The Heidelberg group is mainly focused on the development of transceiver chips and on the building on a demonstrator, whilst the Uppsala group is more focused on the development , construction, and testing of antennas. A study of antennas for 60 GHz data transfer in tracking detectors has been presented in a thesis. In Uppsala the group collaborates on mm-wave technologies with the division of Solid State Electronics. All have expertise in the building and operation of large detectors as well as in the electronic engineering.

- In the context of neutrino detectors, physicists and engineers from the multidisciplinary science and engineering research centre of Argonne work alongside experts from industry. They are specialists of neutrino physics detectors but also have experience in HEP with contributions to the electronics of ATLAS and to the CALICE Collaboration. Not only they explore the use of wireless techniques for detector readout but also for the power transmission, and have demonstrated the feasibility in that context.

More recently, a senior engineer-physicist from CEA-Saclay, having a large experience in HEP experiments such as UA1 at the SPS, ALEPH at LEP, CMS at LHC, had independently from the former groups the idea of using wireless techniques in the context of future large colliders experiments. Becoming aware of the above quoted works, she proposed to those groups and to the most experimented engineers of CEA-Leti to form a Collaboration, joined by a senior physicist of The Corean University of Gangneung University. The CEA engineers have top expertise in all wireless techniques. The CEA-Leti possesses complete CAD design environment and test facilities for test and characterization of chips and antennas. They can offer early access to wireless technologies, and are able to address manufacturability. They have all competencies to adapt wireless techniques for the HEP requirements and to solve technical issues attached to this context, to realise prototypes, and ultimately produce all specifications for mass production at the best cost. The CEA group envisage to complement the 60 GHz developments with studies at higher frequencies in the 240-280 GHz range, which present a great interest for their larger bandwidth and for the smaller size of the electronic circuits and antennas which scales with frequency/wavelength.

The present Collaboration is expected to grow in:

- Manpower:
  - PhD students, post-docs.
  - Application engineers with various competency in electronics design, chipset design, and millimetre wave radio communication systems.
- Laboratory equipment
  - Fully equipped labs to develop and test technology at 60 GHz or above.
- Resources for manpower, equipment, conferences, seminars, travels, training

It will be the task of each working group to estimate the budget required for the R&D in the group, in order to have a global budget estimate for the whole project and specific applications.
Some milestones need to be defined. The first milestone might be the completion of a demonstrator over a given time period. The necessary budget has to be carefully evaluated and regularly reviewed with the project progress.

## 8. Conclusions

In the last decade, there has been a growing interest for wireless techniques and the domain is advancing faster and faster. Considerable efforts are deployed in terms of performances (spectral efficiency, link reliability), miniaturisation, electric power consumption and manufacturing cost. Therefore we believe that these techniques will be widely used in the future, and would be of benefit to particle physics in the medium to long-term future. Furthermore, these developments could also benefit to other industrial, medical and scientific applications. Thus we propose a collaboration to make the necessary step to deploy these techniques in the particle-physics field by defining a common platform for research and development to optimize effectiveness and cost, with the aim of designing and testing wireless demonstrators for large instrumentation systems.

## Acknowledgments

The authors wish to thank Fritz Caspers for challenging discussions.

# Definition of Acronyms

| | |
|---|---|
| **3GPP** | 3$^{rd}$ Generation Partnership Project |
| **BiCMOS** | Integration of Bipolar junction transistor and Complementary Metal-Oxide-semiconducteur in a single circuit |
| **CA** | Channel Access |
| **CAD** | Computer-Aided Design |
| **CSMA** | Carrier Sense Multiple Access |
| **EMC** | Electromagnetic Compatibility |
| **EMI** | ElectroMagnetic Interference |
| **FTE** | Full Time Equivalent |
| **HBT** | Heterojunction Bipolar Transistor |
| **ISM** | Industrial Scientific Medical |
| **LOS** | Line-Of-Sight |
| **LTE** | Long Term Evolution (4G, LTE-B refers to 5G) |
| **MIMO** | Multiple Input – Multiple Output |
| **MPW** | Multi Project Wafer |
| **OAM** | Orbital Angular Momentum |
| **OFDM** | Orthogonal Frequency Division Multiplex |
| **OOK** | On-Off Keying |
| **OSI** | Open Systems Interconnection |
| **PMT** | Photo-Multiplier Tube |
| **QAM** | Quadrature Amplitude Modulation ($2^N$QAM refers to Nb of Information/symbol, then 4096QAM refers to 12 b) |
| **RF** | Radio Frequency |
| **SCT** | Semi-Conductor Tracker (ATLAS) |
| **SDMA** | Space-Division Multiple Access |
| **SHARP** | Stationary High Altitude Relay Platform |
| **SPS** | Solar Power generating Satellites |
| **SWIPT** | Simultaneous Wireless Information and Power Transfer |
| **UMTS** | Universal Mobile Telecommunication System |
| **UNII** | Unlicensed National Information Infrastructure |
| **UWB** | Ultra Wide Band |
| **WiMAX** | Worldwide Interoperability for Microwave Access |
| **WIT** | Wireless Information Transmission |
| **WLAN** | Wireless Local Area Network |
| **WPT** | Wireless Power Transmission |

**WSN** Wireless Sensor Network

# References

[1] "Planning the Future of U.S. Particle Physics (Snowmass 2013), Chapter 8: Instrumentation Frontier, FERMILAB-CONF-14-019-CH08.
[2] R. Brenner and S. Cheng, "Multigigabit wireless transfer of trigger data through millimetre wave technology", 2010 *JINST* 5 C07002.
[3] D. Pelikan et al., "Wireless data transfer with mm-waves for future tracking detectors", 2014 *JINST* 9 C11008.
[4] H.K. Soltveit et al., "Multi-gigabit wireless data transfer at 60 GHz", 2012 *JINST* 7 C12016.
[6] S. Dittmeier *et al.*, "60 GHz wireless data transfer for tracker readout systems—first studies and results", 2014 *JINST* 9 C11002.
[7] H. K. Soltveit et.al., "Towards Multi-Gigabit readout for the ATLAS silicon microstrip detector", Nuclear Science Symposium and Medical Imaging Conference (NSS/MIC), 2013 IEEE.
[8] T.S. Rappaport et al., "State of the Art in 60-GHz Integrated Circuits and Systems forWireless Communications", Proceedings of the IEEE, Vol. 99, No. 8, August 2011.
[9] A. Siligaris et. al."A 60 GHz Power Amplifier With 14.5 dBm Saturation Power and 25% Peak PAE in CMOS 65 nm SOI," IEEE Journal of Solid-State Circuits (JSSC), vol.45, no.7, p.1286-1294, July 2010.
[10] A. Siligaris et al., "A 65 nm CMOS fully integrated transceiver module for 60 GHz Wireless HD applications," International Solid-State Circuits Conference (ISSCC), San Francisco, 20-24 February 2011.
[11] A. Siligaris et al., "A 65-nm CMOS fully integrated transceiver module for 60-GHz Wireless HD applications", IEEE Journal of Solid-State Circuits (JSSC), December 2011.
[12] H. Kaouach et al., "Wideband low-loss linear and circular polarization transmit-arrays in V-band," IEEE Transactions on Antennas and Propagation, July 2011.
[13] U.R. Pfeiffer and D. Goren, "A 20 dBm fully-integrated 60 GHz SiGe power amplifier with automatic level control", IEEE Journal of Solid-State Circuits, vol. 42, no. 7, p. 1455-1463, 2007.
[14] S. Reynolds et al., "A silicon 60-GHz receiver and transmitter chipset for broadband communications", IEEE Journal of Solid-State Circuits, vol. 41, p. 2820-2831, December 2006.
[15] Y. Fu et al., "Characterization of integrated antennas at millimeter-wave frequencies", International Journal of Microwave and Wireless Technologies, p. 1-8, 2011.
[16] A. Siligaris et al., "A 60 GHz UWB impulse radio transmitter with integrated antenna in CMOS 65 nm SOI technology", 11th IEEE Topical Meeting on Silicon Monolithic Integrated Circuits in RF Systems, Phoenix, 17-19 January 2011.
[17] L. Dussopt, "Integrated antennas and antenna arrays for millimetre-wave high data-rate communications", 2011 Loughborough Antennas and Propagation Conference (LAPC 2011), 14-15 November 2011, Loughborough, UK.
[18] L. Dussopt, et al., "Silicon Interposer with Integrated Antenna Array for Millimeter-Wave Short-Range Communications", IEEE MTT-S International Microwave Symposium, 17-22 June 2012, Montreal, Canada.
[19] J.A. Zevallos Luna et al., "Hybrid on-chip/in-package integrated antennas for millimeter-wave short-range communications", IEEE Transactions on Antennas and Propagation, vol. 61, no. 11, p. 5377-5384, November 2013.
[20] A. Siligaris, et al., "A low power 60-GHz 2.2-Gbps UWB transceiver with integrated antennas for short range communications", 2013 IEEE RFIC conference, 2-4 June 2013, Seattle, Washington, USA.

[21] Y. Lamy et al., "A compact 3D silicon interposer package with integrated antenna for 60 GHz wireless applications," IEEE International 3D Systems Integration Conference (3DIC), 2-4 October 2013, San Francisco, CA, USA.

[22] J.A.Z. Luna et. al., "A packaged 60 GHz low-power transceiver with integrated antennas for short-range communications", Radio and Wireless Symposium (RWS), 2013 IEEE , p. 355-357, 20-23 January 2013.

[23] J.A.Z. Luna et al., "A V-band Switched-Beam Transmit-array antenna," to appear in International Journal on Microwave and Wireless Technology , 2014.

[24] J. Antes et al., "System concept and implementation of a mmW wireless link providing data rates up to 25 Gb/s" , presented at the IEEE COMCAS, Tel Aviv, Israel, 2011.

[25] J.M. Guerra et al., "A 283 GHz low power heterodyne receiver with on-chip local oscillator in 65 nm CMOS process", Radio Frequency Integrated Circuits Symposium (RFIC), 2013 IEEE, p. 301-304, 2-4 June 2013.

[26] G. Fettweis et al., "Entering the Path Towards Terabit/s Wireless Links", Proceedings of the IEEE European Conference on Antennas and Propagation (EUCAP), 2011.

[27] "ISSCC 2013 Trends", International Solid State Conference 2013.

[28] "ISSCC 2014 Trends", International Solid State Conference 2014.

[29] Yan Yan et al., "High-capacity millimetre-wave communications with orbital angular momentum multiplexing" Nature Communications 5, Article number: 4876, doi:10.1038/ncomms5876, Published 16 September 2014.

[30] W. C. Brown, "The History of Wireless Power Transmission", *Solar Energy* Vol. 56, No. 1, p. 3-21, 1996.

[31] Spectrum.ieee.org/solarspace0514

[32] P. De Lurgio et al., "A Prototype of Wireless Power and Data Acquisition System fol Large Detectors", arxiv/1310.1098 [physics.ins-det].

[33] P. Ferrari, G.Giorgi, C. Narduzzi, S. Rinaldi, IEEE Transactions on Instrumentation and Measurement, Vol.63, No.11, November 2014.